\documentclass{article}

\usepackage{arxiv}

\usepackage[utf8]{inputenc} 
\usepackage[T1]{fontenc}    
\usepackage{hyperref}       
\usepackage{url}            
\usepackage{booktabs}       
\usepackage{amsfonts}       
\usepackage{nicefrac}       
\usepackage{microtype}      
\usepackage{lipsum}
\usepackage{graphicx}
\graphicspath{ {./images/} }
\usepackage{graphicx,wrapfig,lipsum}

\usepackage{amsmath}
\usepackage{lineno}
\usepackage{float}

\usepackage{graphicx}
\usepackage{caption}
\usepackage{subcaption}

\usepackage{textcomp}
\usepackage{gensymb}
\usepackage{comment}

\title{Simulation of a first case study for magnetic field imaging with the Magic-µ technique}

\author{
 Hamid Basiri \\
 Interdisciplinary Graduate School of Engineering Science\\
 Kyushu University\\
 6-1 Kasuga-koen, Kasuga-shi, Fukuoka, 816-8580, Japan\\
  \texttt{basiri.hamid@kyudai.jp} \\
   \And
 Tadahiro Kin \\
 Interdisciplinary Graduate School of Engineering Science\\
 Kyushu University\\
 6-1 Kasuga-koen, Kasuga-shi, Fukuoka, 816-8580, Japan\\
  \And
 Naoya Okamoto\\
 Interdisciplinary Graduate School of Engineering Science\\
 Kyushu University\\
 6-1 Kasuga-koen, Kasuga-shi, Fukuoka, 816-8580, Japan\\
     \And
Eduardo Cortina Gil\\
Centre for Cosmology, Particle Physics and Phenomenology\\
Universit\'e catholique de Louvain\\
Chemin du Cyclotron 2, B-1348 Louvain-la-Neuve, Belgium\\
    \And
Andrea Giammanco\\
Centre for Cosmology, Particle Physics and Phenomenology\\
Universit\'e catholique de Louvain\\
Chemin du Cyclotron 2, B-1348 Louvain-la-Neuve, Belgium\\
}
\begin{document}
\maketitle
\begin{abstract}
So far, most of the developments in muography (or cosmic-ray muon radiography) have been based on either the scattering or the absorption of cosmic-ray muons produced by the nuclear interactions between primary cosmic-rays and the nuclei of the Earth's atmosphere. Applications of muography are increasing in various disciplines. A new use of this technique to measure a magnetic field has recently been proposed by our group. This new application takes advantage of the electric charge of cosmic-ray muons, which causes them to change their trajectory due to the Lorentz force generated by a magnetic field. In this study, we present a feasibility study of the proposed technique by simulating a simple dipole magnet using the three-dimensional finite element solution package AMaze, together with the PHITS Monte Carlo simulation tools. The distribution of magnetic field flux densities around the magnet was calculated in AMaze and entered into the PHITS code. Positive and negative cosmic-ray muons were generated based on the PHITS-based analytical radiation model (PARMA). A comparison of the count rate maps of the detected muons on two position-sensitive scintillator detectors for the magnetic field ON and OFF was studied using PHITS. The simulation results show the effect of the magnet on the count rate maps and are promising for the newly proposed application of cosmic-ray muons, the imaging of a magnetic field.
\end{abstract}


\section{Introduction}

Muons are elementary particles similar to the electron and belong to the lepton group of elementary particles. Their mass is about 207 times greater than that of the electron, and the average energy of muons reaching the Earth's surface (with a flux of about 1 muon per square centimeter per minute) is about 4 GeV. These particles are continuously produced in the upper layer of the atmosphere by the nuclear interactions of high-energy primary cosmic-rays with atmospheric nuclei. Muons are unstable and decay into electrons and neutrinos with an average lifetime of 2.2$\micro s$. However, because of the large time-dilation effect due to their high energies, they can reach the Earth's surface and, depending on their energy, can penetrate large thicknesses of material. The particular mass of muons (intermediate between an electron and a proton) causes a negligible probability of producing electromagnetic cascades and relatively low energy losses due to ionization, two properties that together with the absence of strong nuclear interactions make them a good tool for imaging large-scale structures on Earth. To exploit this natural atmospheric muon flux, the technique of muography (or muon radiography/tomography) was developed in a similar way to X-ray radiography \cite{bonechi2020atmospheric}. 
This technique has been used for various applications, from the study of hidden chambers in pyramids \cite{alvarez1970search} to the inspection of the density distribution or internal structure of objects such as volcanoes \cite{tanaka2019japanese} and nuclear reactors \cite{miyadera2013imaging}, taking into account the absorption of muons in the volume of interest. Following recent developments in the field of detectors, research groups have attempted to build various portable and small muography detectors (see e.g. \cite{wuyckens2019portable}). This has led to thinking about new applications of muography, such as using muons as a tool for infrastructure inspection \cite{chaiwongkhot2018development}. In addition to the above applications, cosmic-ray muons have been introduced as a candidate for homeland security and national border inspections by considering the multiple Coulomb scattering of muons in matter, especially in high-Z materials such as nuclear materials \cite{borozdin2003radiographic}. \par

While various companies and research groups around the world are trying to develop applications and technologies for muography to exploit this free, hazard-free, and available source, our team is investigating a new application for cosmic-ray muons: imaging a magnetic field, based on the deflection of cosmic-ray muons due to the Lorentz force in a magnetic field. The project is called ``Magic-$\mu$'', an acronym for MAGnetic field Imaging by Cosmic-ray MUons. This new technique takes advantage of the charge on cosmic muons that causes them to change their trajectory after passing through a magnetic field. 

One possible application for this new method could be the monitoring of variations in the magnetic field used in nuclear fusion reactors, where standard magnetic measurement methods are challenging due to the special conditions of extremely high magnetic field and high neutron flux (the latter could cause degradation of the superconducting coil). 
The general principles and overall strategy are presented in~\cite{Tadahiro-these-procs}, while in this study, the feasibility of the Magic-$\mu$ technique was investigated by simulating the measurement of cosmic muons traversing a magnetic field generated by a permanent magnet.

\section{Material and Methods}

In this study, we used the three-dimensional finite element solution package AMaze and the Monte Carlo code PHITS (Particle and Heavy-Ion Transport code System) \cite{sato2018features,humphries2015electric}. In the first step, the magnetic flux density of a simple dipole magnet was simulated in the AMaze software. The dimension of each magnet plate was considered to be $20\text{ cm} \times 20 \text{ cm} \times 1.1 \text{ cm}$ and the gap between two magnet plates was 9~cm. The plates were surrounded by a yoke of a thickness of 1~cm. A quite uniform magnetic flux density of 0.2~T was created in the gap region and the 3D distribution of magnetic flux densities was calculated in a $40\text{ cm} \times 40 \text{ cm} \times 40 \text{ cm}$ voxelized region around the magnet. \par

The next step was to convert the calculated distribution of magnetic flux densities into a format readable by PHITS. Then, the geometry of the magnet was defined in PHITS and the cosmic-ray muons were generated as originating from a hemisphere with a radius of 560 cm, assuming the local geomagnetic cut-off corresponding to the location of the Chikushi campus of Kyushu University ($33 \degree 31^{\prime} N,$ $13 \degree 28^{\prime}E$, 35 m above sea level) using the model PHITS-based analytical radiation model (PARMA) implemented in the PHITS code \cite{sato2008development}. The number of events corresponds to 152 hours of data collection and is large enough to achieve a statistical uncertainty of 1\% in the region of interest. 

In this preliminary study, we simulate a minimalistic detector set-up, consisting only of two position-sensitive plastic scintillator detectors (mu-PSD). 
More powerful imaging would be possible with additional mu-PSD, as foreseen in the full Magic-mu methodology, for example upstream of the magnetic region (to follow the trajectory before and after the deflection) and at different angles (for 3D reconstruction of the magnetic field). 
However, we start with this simplified configuration as it will give a first useful indication of the effect on muon imaging of the interplay of the main physical effects: muon absorption and scattering in the materials, and magnetic deflection. 

To observe the influence of the magnet on traditional muography imaging, we placed a lead block with the height of 30~cm, 5 meters above the first mu-PSD. The position of the lead block was determined in a way that it represented a single pixel in the center of the count rate map considering the detector resolution. The lead block is composed of smaller lead blocks of the dimension of $20\text{ cm} \times 10 \text{ cm} \times 5 \text{ cm}$, disposed in such a way that we maximize the absorption of muons by lead while not exceeding the maximum load limit of the roof of the building, which is 300 kg/$m^2$. The distance between the two mu-PSDs is 31.9~cm and each mu-PSD has a size of $29.6\text{ cm} \times 29.6 \text{ cm}\times 0.4 \text{ cm}$ with 16 channels for each x and y direction, leading to a pixel area of $3.4 \text{ cm}^2$. The geometry and dimensions used in this simulation can be found in Figure \ref{fig:phits}.

\section{Results and Discussion}

\begin{figure}[ht]
\begin{subfigure}{0.33\textwidth}
\includegraphics[width=\linewidth]{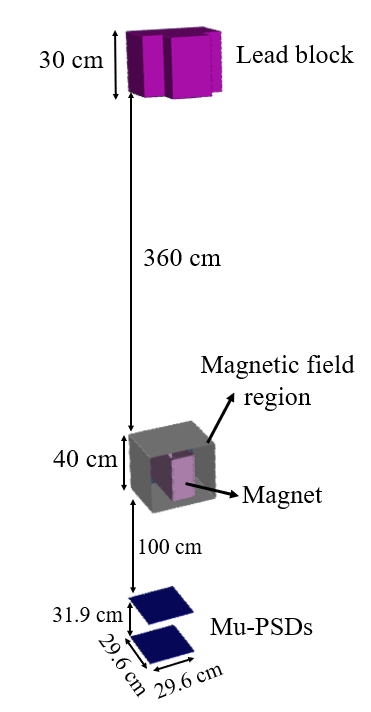}
    \caption{Geometry of simulation in PHITS.}
     \label{fig:phits}
\end{subfigure}
\hspace*{\fill}
\begin{minipage}{0.64\textwidth}
\begin{subfigure}{\linewidth}
\includegraphics[width=\linewidth,height=2in]{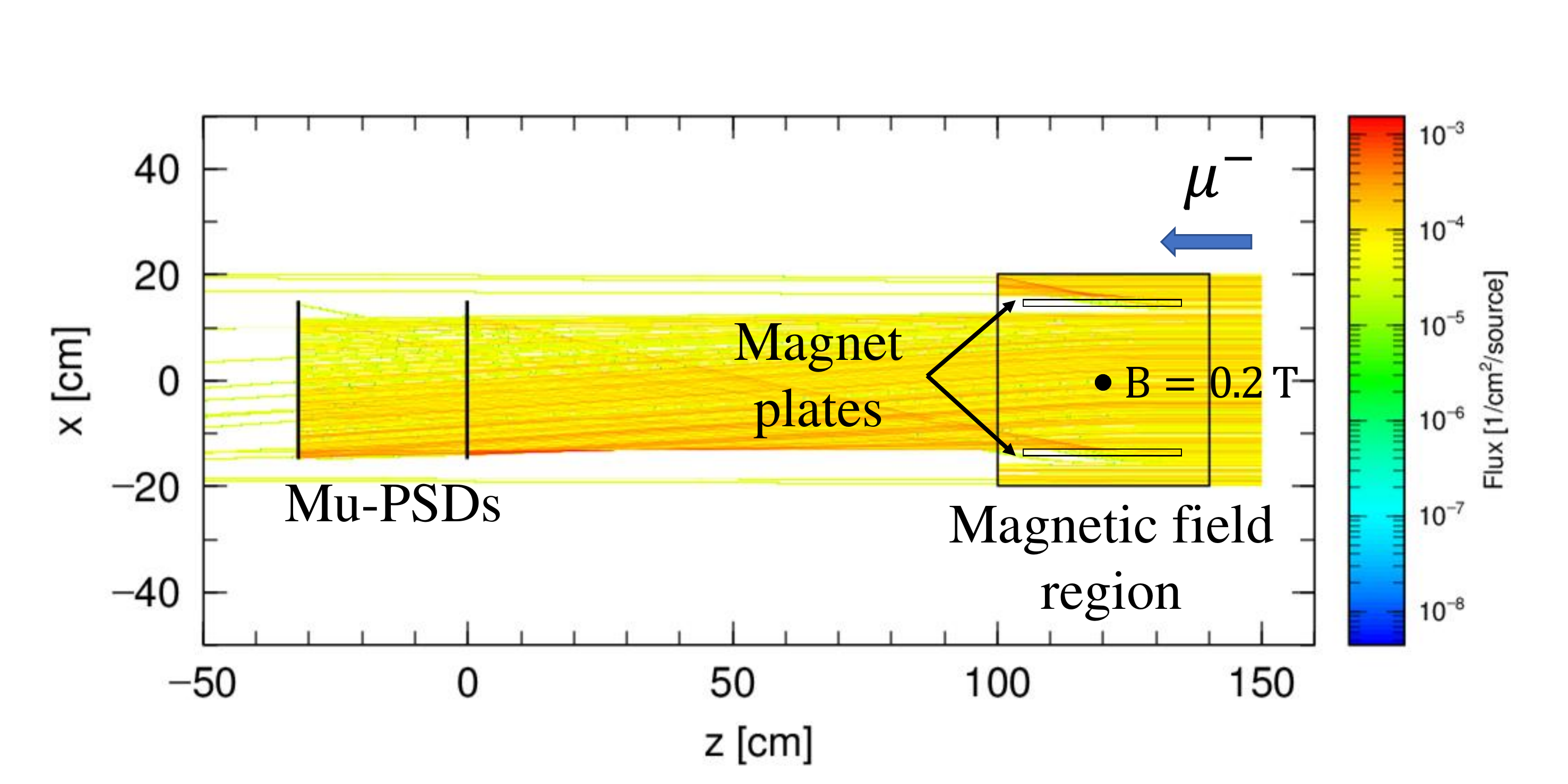}
         \caption{Negative muons deflection due to magnet.}
              \label{fig:neg}
\end{subfigure}

\begin{subfigure}{\linewidth}
\includegraphics[width=\linewidth,height=2in]{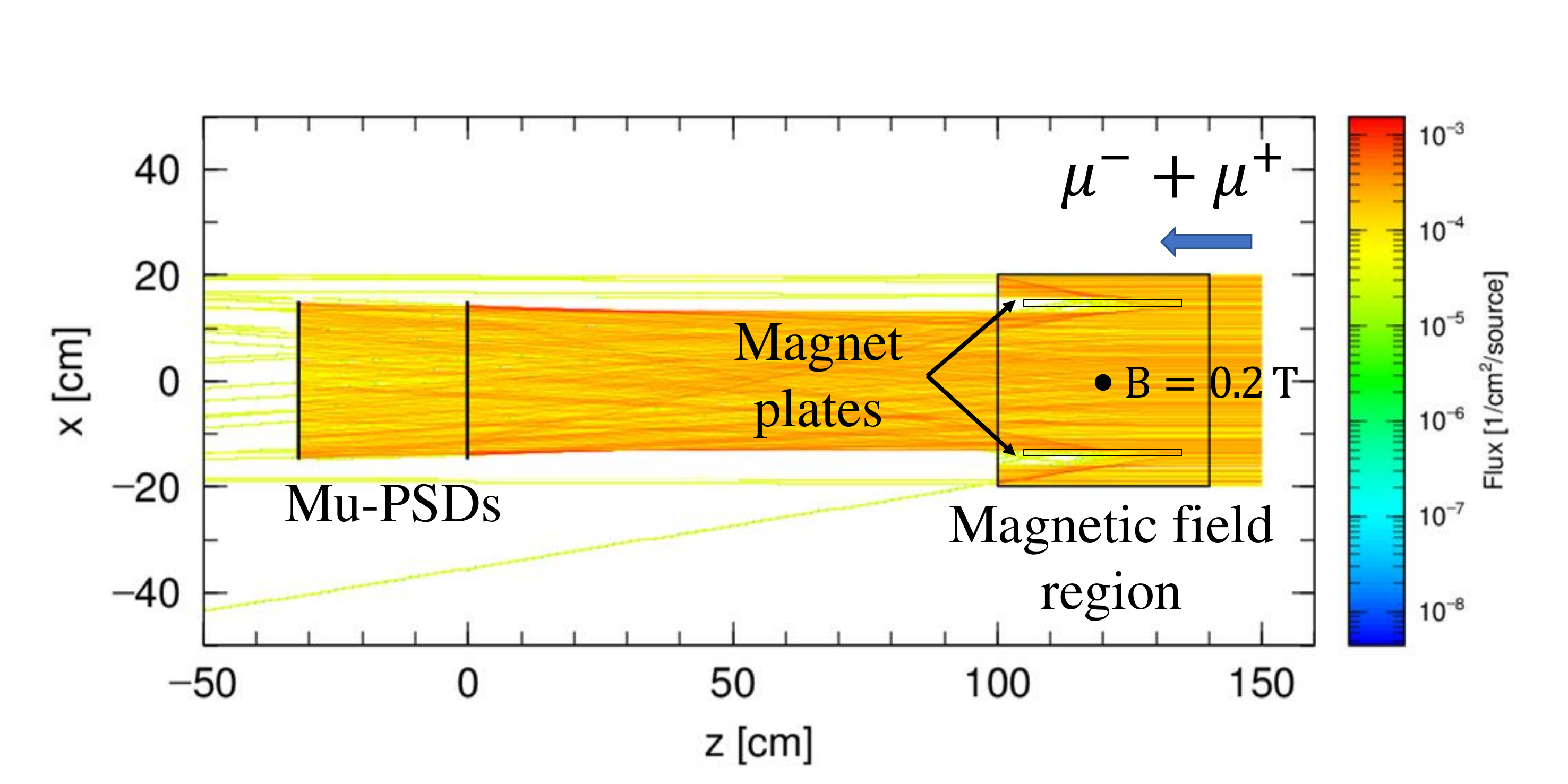}
         \caption{Positive and negative muons deflection due to magnet.}
         \label{fig:both}
\end{subfigure}
\end{minipage}

\caption{Simulation of the set-up for this feasibility study (a), and effect of the magnetic field on the muon trajectories (b,c) illustrated for the case of 100~MeV energy.}
\end{figure}

To illustrate the defined magnetic field region and verify the conversion of AMaze to PHITS, we used a flat source with 100~MeV muons. As shown in Figures \ref{fig:neg} and \ref{fig:both} , the deflection of the muons is proportional to the magnitude of the magnetic field flux density in the defined meshes. The deflection is almost zero at the entrance of the magnetic field region and maximum in the middle. Moreover, positive and negative muons are deflected in opposite directions as expected. Due to the complexity of the magnetic field near the edge of the magnet, some muons are deflected in the opposite direction. The empty area in the magnetic field region corresponds to the magnet plates where these 100~MeV muons were absorbed.

The information about the position, direction and energy of the muons hitting the first mu-PSD was analyzed using the ROOT framework~\cite{brun1997root}. We simulated three different conditions:\\
(a) Open sky: only the mu-PSDs in the PARMA source.\\
(b) Background: lead block and magnet while the magnetic field region was turned OFF.\\
(c) Foreground: lead block and magnetic field ON.\\ \par

\begin{figure} [H]
     \centering
     \begin{subfigure}[b]{0.33\textwidth}
         \centering
         \includegraphics[width=\textwidth]{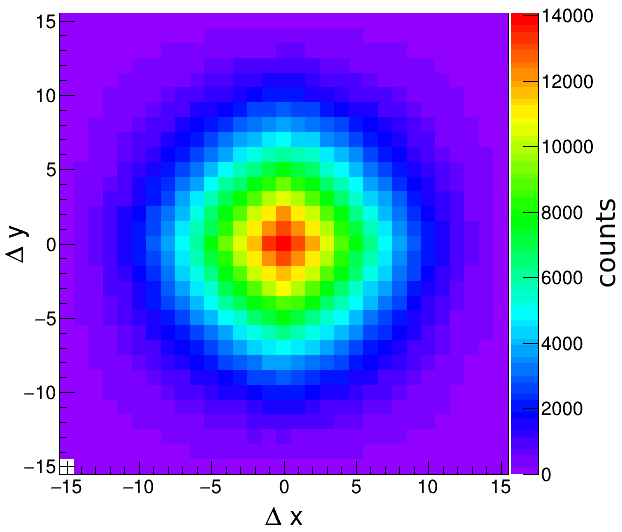}
         \caption{Open sky}
         \label{fig:Open}
     \end{subfigure}
     \hfill
     \begin{subfigure}[b]{0.33\textwidth}
         \centering
         \includegraphics[width=\textwidth]{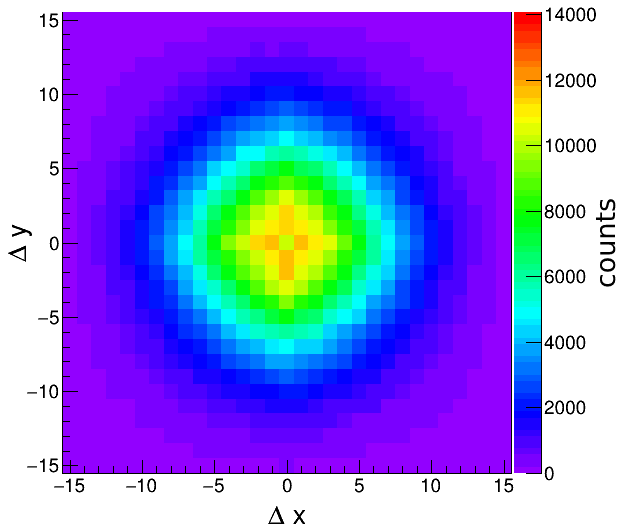}
         \caption{Background}
         \label{fig:Bg}
     \end{subfigure}
     \hfill
     \begin{subfigure}[b]{0.33\textwidth}
         \centering
         \includegraphics[width=\textwidth]{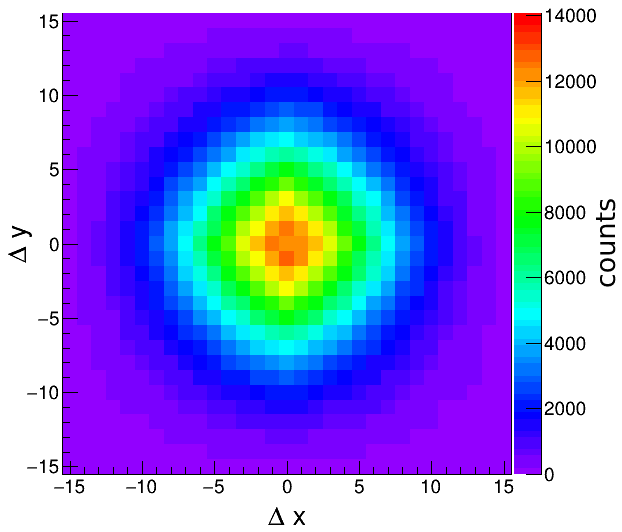}
         \caption{Foreground}
         \label{fig:Fg}
     \end{subfigure}
        \caption{Muon counting rate maps for a realistic simulation of the muon flux.}
        \label{fig:count}
\end{figure}

To study the effect of the magnet, the contrast of the count rate maps for the background and foreground measurements is important, therefore the counting rate maps (muography images of the lead block) for all three conditions were plotted. Figure~\ref{fig:count} shows the vector plots corresponding to the displacement of muons between the upper and lower mu-PSDs for the three stages of this study, where the axes show the difference in the number of pixels for mu-PSD 1 and mu-PSD-2 in the x and y directions ($\Delta x = \Delta y = 0$ corresponds to the vertical muons). As we expected based on the theory and the known distribution of cosmic muons (that approximately follows a $cos^2 \theta$ dependence on the zenith angle), in the open sky measurement we see the maximum flux in the center due to vertical events, while in the background the effect of absorption of muons in the lead block has a large impact on the flux in the center of the plot. In the foreground simulation, the distribution of counts has changed due to the effect of the magnetic field, and the central pixel corresponding to the image of the lead block is blurred. The increase in the other pixels is due to the deflection of the vertical events. In addition, some muons that could not reach the detector during the background measurement hit the detector after crossing the magnetic field region.


The statistical significance between background and foreground was analyzed using the figure of merit (FOM) \cite{zaitseva2011pulse} defined as equation~\ref{eq1}:

\begin{equation} \label{eq1}
FOM = \frac{S}{FWHM_{FG} + FWHM_{BG}}
\end{equation}

Where $S$ is defined as the difference between the foreground and background counting rates, and FWHM (the full width at half maximum) can be related to the statistical uncertainty $\sigma$ using the equation~\ref{eq2}:
 
\begin{equation} \label{eq2}
FWHM = 2\sigma \sqrt{2 \ln{2}}
\end{equation}

We apply a threshold of $FOM > 2$. Each bin in the counting rate map of the background measurement was compared with the same bin in the foreground measurement and if the calculated FOM value was greater than 2 we considered that bin as magnetic field region. As the result of this analysis a square area in the center of the counting rate map was estimated as a possible magnetic field area. The magnet itself has a rectangular shape, but the magnetic field was present in a cubic volume around the magnet, so the detected region has a square shape.

\section{Conclusion}
For a feasibility study of the proposed magnetic field muography technique, a simulation study was carried out in this study. First, the magnetic field flux density distribution around a  simple dipole magnet (which had a maximum value of 0.2~T) was calculated using the AMaze software. Then, the magnet and magnetic field region were implemented in PHITS and the effect of this magnet on the count rate maps (muography image of a lead block) for the magnetic field ON and OFF status were studied. The results showed that this magnet has a significant impact on the count rate maps and can blur the lead block muography image. This effect can be used for magnetic field analysis. The simulation results are promising for the newly proposed application of cosmic muons, the imaging of the magnetic field. In the near future, we will perform a measurement test and develop an analysis method to estimate the presence of the magnetic field and its strength. Also, we will develop a new approach based on sandwiching the magnet between mu-PSDs.

\bibliographystyle{ieeetr}
\bibliography{references}

\end{document}